\documentclass[runningheads]{llncs}
\usepackage{graphicx}
\usepackage[T1]{fontenc}
\usepackage[utf8]{inputenc}
\usepackage{hyphenat}
\begin{document}
\title{Towards Heuristics for Supporting the Validation of Code Smells}
\author{Luiz Felipi Junionello\inst{1}\and
Rafael de Mello\inst{1}\orcidID{0000-0002-9877-3946}} %\and
%Anderson Uchoa\inst{3}\orcidID{2222--3333-4444-5555}}
%
\authorrunning{Junionello and de Mello}
\institute{CEFET/RJ, Rio de Janeiro, Brazil \\%\and
\email{luiz.junionello@aluno.cefet-rj.br, rafael.mello@cefet-rj.br}}
\maketitle              % typeset the header of the contribution
\begin{abstract}
The identification of code smells is largely recognized as a subjective task. Consequently, the automated detection tools available are insufficient to deal with the whole subjectivity involved in the task, requiring human validation. However, developers may follow different but complementary perspectives for manually validating the same code smell. Based on this scenario, our research aims at characterizing a comprehensive and optimized set of heuristics for guiding developers to validate the incidence of code smells reported by automated detection tools. For this purpose, we conducted an empirical study with 12 experienced software developers. In this study, we invited developers to individually validate the incidence of code smells in 24 code snippets from open-source Java projects. For each validation, developers should provide arguments for supporting their decisions. The study findings revealed that developers tend to look from different perspectives even when they agree about the incidence of a code smell. After coding the 303 arguments given into heuristics and refining them, we composed an optimized set of validation items for guiding developers on manually validating the incidence of eight types of code smells: data class, god class, speculative generality, middle man, refused bequest, primitive obsession, long parameter list, and feature envy. We are currently planning a survey with specialists for identifying opportunities for evolving the set of validation items proposed. 

\end{abstract}
%
%Software engineering is composed of tasks influenced by human aspects~\cite{lenberg2015behavioral}.
%

\section{Introduction} \label{sec:introduction}

The variety of maintenance requests over different source code elements frequently challenge software developers \cite{bennett2000software}. This challenge typically results from the structural complexity of the source code, requiring considerable reading and comprehension efforts from these professionals for performing even simple tasks. For mitigating these efforts, one key practice addresses continuously identifying and combating the incidence of code smells. Code smells are known as indicators of deeper problems within a source code, commonly introduced due to the negligence of good programming practices. The incidence of code smells harms maintenance activities\cite{palomba2018diffuseness}~\cite{lima2020understanding} once it hampers the source code readability and comprehension. Besides, different works associate the incidence of code smells with the acceleration of the software degradation in the long term\cite{van2002design}~\cite{fowler2018refactoring}.

Due to its granularity, manually identifying and fixing smelly code in entire software systems or modules is definitively not a trivial task. In this sense, several detection tools have been proposed in the last decade to automate the detection of code smells \cite{fernandes2016review}. Even though these tools may save effort on identification activities, they cannot be considered the final word ~\cite{paiva2017evaluation}~\cite{fernandes2016review}. After a tool reports candidates to code smell, developers should manually validate these issues, distinguishing false positives from the actual ones. Among others, neglecting such validation may lead developers to unnecessarily modifying several code elements associated with false positives. Consequently, a lack of validating candidates to code smells may result in a considerable waste of maintenance effort. Besides, it may lead developers to accidentally introduce new and even worse issues in the source code, including bugs and design problems~\cite{de2019CHASE}.

However, one can see that manually validating the incidence of code smells is a considerably subjective task. Distant from the generic definitions available at catalogues and raw thresholds, the decision about the incidence of code smells may be highly influenced by contextual factors, including technological, organizational, and human ones~\cite{de2017influence}~\cite{de2017towards}~\cite{de2019CHASE}. Indeed, even colleagues reviewing the same system module may have different interpretations about the incidence of smelly code due to their background ~\cite{de2017influence}~\cite{hozano2018you}. Consequently, the subjectivity involved in the identification of code smells frequently leads developers to disagree on their final decisions\cite{hozano2018you}. On the other hand, the diversity of perspectives from two or more reviewers working together- when possible -contributes to increasing the performance of smell identification tasks ~\cite{de2017influence}~\cite{oliveira2017collaborative}~\cite{oliveira2020collaborative}. However, allocating developers for conducting group reviews may be unfeasible due to several reasons, including schedule and budget restrictions. 

Besides, our recent studies on the developers' social representations~\cite{de2019CHASE}~\cite{de2019ESEM} reveal the need for guiding these professionals on reflecting about several technical and non-technical aspects surrounding the manual validation of code smells. In this way, we argue that coding and reusing heuristics followed by different developers is a promising strategy towards reaching this guidance. By heuristics, we understand a set of particular attention points for enabling the developers' analysis and, consequently, supporting their decision-making. The set of heuristics to be applied in each issue will depend on the type of code smell reported by the detection tool. For instance, let's suppose that a certain tool had detected a possible incidence of the speculative generality smell. This code smell is characterized by a code element written just for accommodating future features, which suggests that it would be discarded. While one reviewer would opt by analysing whether the class can perform some relevant task, a second reviewer would opt by checking whether this class actually needs to be used by other code elements. Once both strategies may be considered valid, a third reviewer may use them as a starting point to reflect before accepting or discarding the reported smell.

% we identified the opportunity of supporting developers to reflect about the incidence of code smells through the use of heuristics. Since It's up to the individual giving the final word, we are proposing heuristics to guide developers in their personal decision making process of validating the reports of smell detection tools. 
%as make that process faster while keeping the programmer's subjectivity aspect.

In this paper, we report our first study towards a comprehensive and optimized set of heuristics for guiding developers on validating the incidence of code smells. This research proposes combining the power of detection of already existing automated tools with the empowered rationale of experienced software developers for improving the effectiveness of smell identification tasks. For composing the first version of this set of heuristics, we conducted a controlled study in which we invited 12 developers to validate 24 suspected code smells reported by a detection tool. These issues address eight popular types of code smells, including \textit{God Class}, \textit{Long Method}, \textit{Long Parameter list}, and \textit{Refused Bequest}. We extracted the projects and code elements used in our study among the dataset used by Hozano et al. for assessing the developers' agreement about the incidence of code smells \cite{hozano2018you}. In the original study, these 24 cases were identified as examples of considerable disagreement among developers. 

In our study, the participants provided arguments for accepting or rejecting the incidence of code smells in each case analysed. In total, 288 arguments were gathered and submitted to open coding, resulting in 40 distinct heuristics distributed among the eight types of code smells investigated. Most of these heuristics go beyond the classical definition of code smells investigated. Then, we used them for compiling a set of 22 validation items (questions) for supporting developers in reflecting on the incidence of the different types of code smells investigated. 

Section~\ref{sec:relatedwork} presents the related work. Section~\ref{sec:studydesign} describes the settings of our empirical study. Section~\ref{sec:results} presents the results of our study, reporting its main findings. It also presents a discussion of the heuristics found for each type of code smell. Section~\ref{sec:heurstics} presents the validation items composed based on the heuristics found. Section~\ref{sec:threats} discuss the main threats to validity identified in our study. Finally, Section~\ref{sec:conclusion} concludes the paper and indicates future work.

%Most of these \textcolor{blue}{The findings of our study reveal that the participants commonly use distinct arguments to accept or reject the incidence of certain code smells. This behavior was specially observed to evaluate the incidence of ............ Most of these arguments address but are limited to the classical definition of the corresponding smells. The Long Parameter List for instance, there were some cases that the argument presented was the fact that it had many complex objects in it's parameters, which isn't commonly mentioned.}

%\textcolor{blue}{Based on these findings we compiled a first set of heuristics to support developers on validating code smells. This set was submmited to the evaluation of eight specialists in code smells about their pertinence and relevance. They were also asked to propose improvements. ....}

\section{Related Work}
\label{sec:relatedwork}

To the best of our knowledge, there was no previous study investigating heuristics to support the manual validation of code smells reported by automated tools. However, there is a set of relevant studies that addressing key motivations for our research. 

%However, we found a recent work using visualization resources to reach a better agreement among developers. Costa~\cite{rossini2020agreement} proposed combining an automated detection based on decision tree learning with a manual validation supported by the visualization of the decision made by the algorithm in the tree. However, after a first empirical evaluation, the authors found low improvement over the developers' agreement.

Regarding the limitations of automated detection tools, Fernandes et al. ~\cite{fernandes2016review} conducted a comprehensive review of automated code smell detection tools. After identifying 84 detection tools and comparing their capabilities and characteristics, the authors selected four tools to perform a in-depth comparison: \textit{inFusion}, \textit{JDeodorant}, \textit{PMD} and \textit{JSpIRIT}. In this comparison, they analyzed the precision and recall of these tools on detecting Long Methods and Large Classes. In general, the tools showed a low performance when detecting different code smells, such as Large Class and Long Method. For instance, although PMD reached 100\% of precision for Large Class, the recall was 14\%. The results for Long Method were better, with JSpIRIT reaching 67\% of recall and 80\% of precision while PMD reached 50\% recall and 100\% of precision.

While the previous study focused on comparing detection tools based on metrics and thresholds, Pecorelli\cite{pecorelli2019comparing} performed a comparative investigation between the performance of smell detection tools based on machine learning and tools based on metric-based heuristics. For this purpose, the authors used a large dataset composed of previously validated code smells. Different from their initial expectation, the authors found that machine learning detection tools are not yet at a stage in which they can be used without manual validation. Among others, the findings from the aforementioned studies indicate the risk involved in only relying on automated smell detection. Depending on the dataset and the smell type, certain tools may reach considerably low levels of precision and recall (coverage). In this way, our research focuses on guiding developers for optimizing their precision on smell identification tasks.

From a human-centered perspective, it is also important to note that the precision of identifying code smells may be influenced by different human aspects \cite{de2017towards}. de Mello et al. \cite{de2017influence} conducted a multi trial empirical study aiming at characterizing the influence of three distinct developers' characteristics over the precision of smell identification tasks: professional experience, familiarity with the module investigated, and the level of interaction among them during the identification tasks (pair x solo). The study findings revealed that developers working in pairs reached higher precision than developers working solo, especially between developers having some professional experience. Besides, the authors found evidence that developers having previous knowledge of the module to review tend to focus on different aspects when identifying code smells from those without this knowledge. Therefore, pairing developers with and without this knowledge represents a better choice once these individuals tend to provide complementary viewpoints of the code elements. More recently, the benefits of collaboratively identifying code smells were also observed over different experiments \cite{oliveira2020collaborative}.

Among the findings of the aforementioned studies, we observed that developers tend to invest considerable effort in taking \textit{ad-hoc} decisions, despite the automated assistance available. Besides, they frequently diverge on their decisions about the same code element, although sometimes showing some concerns in common beyond formal rules and metrics. To better understand this behaviour, we conducted a pioneer investigation on the social representations of the identification of code smells~\cite{de2019CHASE}~\cite{de2019SBES}~\cite{de2019ESEM}. The theory of social representations considers that a task such as identifying code smells is collectively seen based on the set of beliefs, values, and behaviors unconsciously shared between its practitioners~\cite{moscovici1988notes}. Consequently, social representations continuously work as an invisible force influencing how individuals deal with the task. The findings of our investigation on social representations revealed considerable gaps between the research on smell identification and its practice. These gaps suggest the need of building proper support for developers manually validating the incidence of code smells~\cite{de2019ESEM}, reflecting about its semantics, change impact analysis, among others.

The dataset and part of the instrumentation of the study reported in this paper were obtained from an investigation on the agreement among developers on validating automatically detected code smells~\cite{hozano2018you}. Hozano et al. conducted an empirical study in which 75 developers validated the incidence of different types of code smells reported by smell detection tools. The authors found that the level of agreement among developers was low in all types of code smells investigated, ranging from 0.24 to 0.32 (Fleiss Kappa). Also, the authors could not find any relevant difference in the agreement levels when analyzing specific categories of developer's backgrounds. After each round composed of ten evaluations, the developers were asked to summarise the heuristics adopted. As a result, the authors observed that several cases of agreement among developers involved similar heuristics. 

Different from ~\cite{hozano2018you}, our study focus on identifying and analyzing heuristics rather than measuring agreement, assuming that disagreement is intrinsic to the nature of an \textit{ad hoc} identification task. For this purpose, we intentionally selected a subset of the tasks with higher disagreement levels from ~\cite{hozano2018you}, asking developers to justify their decisions after each validation task. In this way, we intend to identify and code the first set of heuristics to lead developers extrapolating their particular insights when validating code smells.

\section{Study Design} 
\label{sec:studydesign}
    
Different developers may adopt different heuristics for concluding that some code element is poorly structured or not. These heuristics typically emerge from an \textit{ah hoc} and individual reasoning process. Consequently, the heuristics adopted by them may significantly vary but may also be limited to single perspectives. By having this in mind, our research aims at characterizing a comprehensive set of heuristics used by developers for validating the incidence of code smells. Based on these heuristics, we intend to compose an optimized set of hands-on validation items to support this task. For this purpose, we understand that extracting a diverse set of arguments from different developers represents one valid strategy. 
    
\subsection{Research Question} 
    
Based on our research goal, we defined the following research question: \\

\textit{Which heuristics do developers follow to validate the incidence of code smells?}\\

By answering this research question, we want to characterize how different developers validate the possible incidence of code smells detected in the context of real software projects. From this, we want to discover more common criteria and possibly unexpected arguments that may help developers in future validation tasks. Based on this knowledge, we will develop more relevant and accurate sets of heuristics for supporting the validation of each type of code smell investigated. 

For this purpose, we designed a controlled experiment partially inspired in the experimental settings of the Hozano et al. \cite{hozano2018you} empirical study. In that work, the authors' goal was to measure the level of agreement among developers about the possible incidence of code smells in 225 different code snippets. These code code snippets were extracted from five popular open-source projects developed using the Java programming language: \textit{GanttProject}, \textit{Apache Xerces}, \textit{ArgoUML}, \textit{jEdit}, and \textit{Eclipse}. From this set, we extracted a subset composed of three code snippets for each type of code smell investigated in our study. All the code snippets selected resulted in considerably low levels of agreement in the original study. Thus, we expect that diverse sets of heuristics may emerge from different on evaluating them. 

\subsection{Population and Sample}

The target population of our study is composed of software developers having knowledge about code smells and code reviews. Considering the findings of previous studies with developers (see Section \ref{sec:relatedwork}) and our research objective, we established the settings of our study sample. First, we opted by investigating developers validating code smells individually once developers would feel free to provide more diverse and unique arguments in this scenario, which is more interesting for reaching our goal. Second, we assured that all study participants have solid professional experience with software development, leading to more reliable insights. Third, considering the nature of the projects involved in the study (large open-source projects), we checked that none of the study participants have previous knowledge of the modules analysed. It would lead developers to feel more comfortable providing less biased arguments, especially those favorable to the incidence of poorly structured code.

The study sample is composed of 12 Master/Doctorate students having solid experience with software development. Besides, most of these professionals are specialists in code smells. After executing the study tasks, we applied a characterization form. In this form, we asked about the experience of the study participants from three distinct perspectives: \textit{self-assessment}, \textit{years of experience}, and \textit{number of projects}. From the self-assessment perspective, most participants declared high or very high experience levels in software development (9/12), as well in Java programming (7/12). Besides, no participant declared having no development experience in Java. From the 12 participants, eight declared experience in identifying code smells. From these, six also conducted researches on this topic. Table \ref{tab:statisticsXP} summarizes the average experience of the participants in terms of years and number of projects on each skill.

\setlength{\tabcolsep}{10pt} % Default value: 6pt
\renewcommand{\arraystretch}{1.1} % Default 

\begin{table}[ht!]
	\centering
	\caption{Average experience of the participants in the different skills measured.}
	\label{tab:statisticsXP}
    \begin{tabular}{|l|c c c|}
    \hline
\textbf{Metric} & \textbf{Software Dev.} & \textbf{Java} & \textbf{Smell Ident.}  \\
\hline
years of experience     & 5.92    & 4.75      & 1.00  \\
\# of projects                & 9.42    & 7.17      & 3.08  \\ \hline
\end{tabular}
\end{table}

The different perspectives used for characterizing the participants' experience led us to conclude that the sample investigated is experienced in building software for different Java projects. Besides, most of the participants are also skilled in the identification of code smells. In the characterization form, we also asked the participants to briefly summarize their experience in software development. Based on the answers provided, we also observed that most of the developers have experience working with different programming languages, building systems for different domains. 

\subsection{Instrumentation}

Through a validation form, we asked the participants to individually validate the incidence of code smells over 24 different code snippets, three to each code smells investigated: \textit{Primitive Obsession, Long Parameter List, God Class, Data Class, Speculative Generality, Feature Envy, Middle Man,} and \textit{Refused Bequest}. All 12 participants evaluated the same code snippets. The code snippets used in the study were picked from the same database used in Hozano et al. study \cite{hozano2018you}. For each validation task, the validation form indicated the exact location of the candidate to code smell in the code snippet. If needed, the study participants also had the option to access the whole source code of each project involved in the study.

After performing each validation task, the subjects were asked to justify their decision, providing detailed arguments on why they concluded that a certain code snippet is smelly or not. For supporting a deeper and accurate analysis, the participants were previously prepared to allocate four hours of their time to perform all the tasks. Besides, each participant received the summarized definition of each type of code smell involved in the study.

%It cParticipants were informed the location of the potential Code  Smell and had access to the full code. Then, to help us answer our RQ we asked them to report on whether said section of the code was smelly or not, and to give detailed explanation on the reasoning behind their answer.
    
\subsection{Data Analysis}
\begin{figure}[ht!]
\centering
\includegraphics[width=200pt]{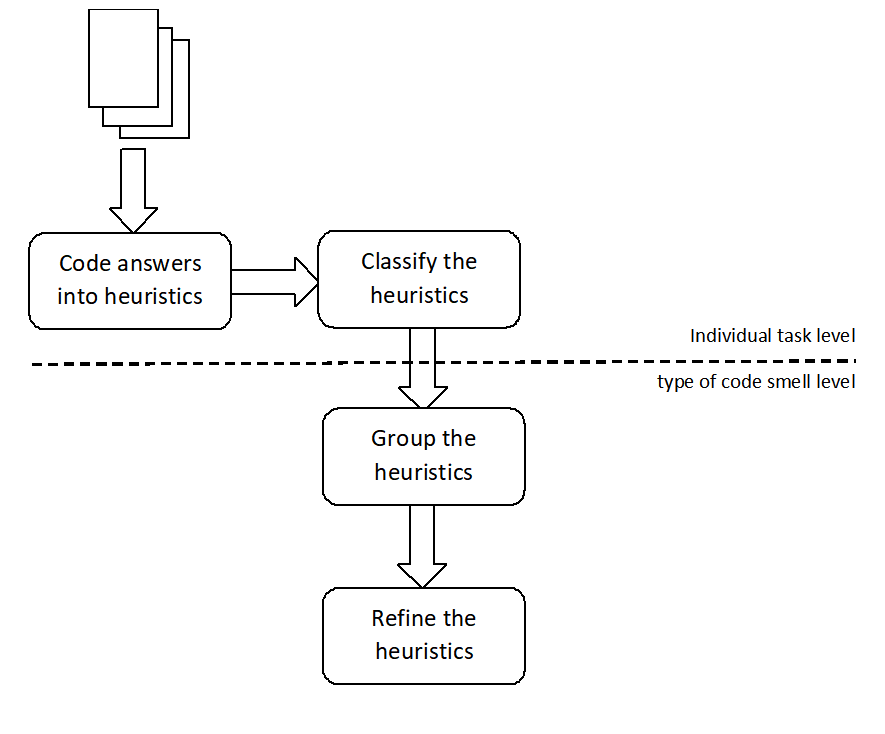}
\caption{The steps followed for coding the heuristics.}
\label{fig:codingprocess}
\end{figure}

We summarize the data analysis procedures in Figure \ref{fig:codingprocess}. We first performed open coding over the arguments given by each participant to each validation task. During the coding process, we tried to identify and categorize the heuristics behind each answer given. In this way, we considered mapping the different actions performed and eventual criteria adopted by the participants for decision making. In the second step, we classified the set of heuristics given on each individual validation as favourable to the acceptance of the code smell (accepting heuristics) or rejecting the code smell (rejecting heuristics). This classification was based on the final decision made by each participant to each task. Then, we grouped all the accepting/rejecting heuristics coded by code smell type. Finally, we performed a refinement of the results (fourth step). For this, we identified opportunities for grouping similar heuristics, eliminating redundancies. We also identified opportunities to splitting too high-level heuristics into two or more.

\section{Results}
\label{sec:results}

The 12 participants performed all the 24 validation tasks, resulting in a set of 288 validations. In total, 301 arguments were coded, including repetitions. From these, 32 (11.80\%) arguments were discarded since they do not address rationale or criteria followed by the participants to support their decisions. In most of these cases, the participants only reported evasive arguments, such as "It has/has not a smell". From the remaining 271 arguments, we found that 57.93\% of them address accepting the code smell while 42.07\% are favorable to rejection. After step 4, 84 distinct arguments were identified considering all the code smells evaluated, 41 for acceptance and 43 for rejection. 

Tables \ref{tab:heuristicsLPL01} exemplifies the coding process by showing the heuristics coded based on the arguments given for the first validation task addressing the incidence of the Long Parameter List code smells. Table \ref{tab:heuristicsLPL} present the final set of heuristics refined for Long Parameter List, considering the heuristics coded from the three validation tasks. 

\begin{table}[ht!]
	\centering
	\caption{Heuristics coded for the first validation task about the incidence of Long Parameter List and their frequency (f).}
	\label{tab:heuristicsLPL01}
    \begin{tabular}{|l c|l c|}
    \hline
 
\textbf{Accepting Heuristics} & \textit{f} & \textbf{Rejecting Heuristics} & \textit{f} \\
\hline
Too many complex parameters & 2 & Needed parameters & 2\\
Too many parameters & 1 & All parameters are used & 2\\
Unused parameters & 1 & Acceptable number of & 2\\
Parameters should be  & 1 & parameters & \\
encapsulated & & It is a constructor & 1 \\\hline
Total &5 & Total &7 \\\hline
\end{tabular}
\end{table}

\begin{table}[ht!]
	\centering
	\caption{Refined set of heuristics addressing the incidence of Long Parameter List and their frequency (f).}
	\label{tab:heuristicsLPL}
    \begin{tabular}{|l c|l c|}
    \hline
\textbf{Accepting Heuristics} & \textbf{f} & \textbf{Rejecting Heuristics} & \textbf{f} \\
\hline
Too many parameters & 6 & Needed parameters & 5\\
Too many complex parameters & 5 & All parameters are used & 5\\
Parameters should be & 3 & It is a builder & 4\\
encapsulated & & Acceptable number  & 2 \\
Unused parameters & 2 & of parameters & \\
Unnecessary parameters & 2 & Easy to understand & 2  \\
Inappropriate use of builder & 2 & & \\\hline
Total &20 & Total &18 \\\hline
\end{tabular}
\end{table}

The following subsections summarize the final set of heuristics coded by type of code smell. Some of these heuristics directly address the general definitions of the code smells. However, we also found several other heuristics provided by the participants extrapolating these definitions. These findings address our initial expectation that developers would apply \textit{ad hoc}, implicit, and useful heuristics for evaluating the incidence of a code smell.
    
\subsection{Data Class}

Most of the arguments given to accept or reject the incidence of Data Classes surround the concern on whether the class was actually used just to store data or whether it has logical methods to process its own data. However, several heuristics were employed. For instance, while most of the arguments focus on the incidence/lack of getters and setters, other ones address the effective role of the class methods and their constructors. Besides, other arguments address to which extent data from the class is externally manipulated.

%, such as the incidence of concrut ones Which is on par with the traditional definition that classes should have logic methods to . The main argument to identify that was on whether or not the class had methods besides getters and setters, sine they would be needed even in the case of it being a Data Class.

\subsection{Feature Envy}
The arguments favorable to the incidence of Feature Envy predominantly address the fact that the code element only accesses external data. Besides, the scope of the responsibilities involved was also taken into account. Different heuristics were employed for rejecting the incidence of this smell in the code snippets analysed. For instance, one heuristic employed was that just a single external object is manipulated. Another heuristic address the importance of the so-called "envy" behaviour as necessary for supporting external objects. Besides, a common heuristic adopted addresses observing the balancing between the manipulation of internal and external data.  
%class  , instance, the stayed in line with the traditional definitions. Most participants were concerned on whether the class relied on retrieving too much external data or not to complete it's tasks, which could indicate the class's lack of data that is often needed. 
    
\subsection{God Class}
        
The number of responsibilities and lines of code were the two more common arguments reported for accepting a God Class. While in some cases the number of lines of code was used as a single argument, in other cases the high amount of lines of code was interpreted as a side effect of assuming several responsibilities. Besides, the participants argued in other cases that even large classes having too many responsibilities are not smelly once their responsibilities seems pertinent.%it was considered that th So we found that class size should only be a factor to validate a God Class when it's when it get in the way of the class's comprehension.

\subsection{Long Parameter List}
In this code smells, developers employed several heuristics. The more common heuristic address reflecting about the amount of the parameters listed. However, other ones address specific characteristics of the code snippets analysed, such as the complexity of the parameters, the possibility of encapsulating them, their relevance, and the role of the corresponding method in the context of eventual design patterns adopted. 
%For instance,  but the ones that stood out were the ones that mentioned the presence of too many complex parameters, which is something not commonly used as a defining factor of this kind of Code Smell.

%presented to be one of the most confusing Smell to the participants, just behind Refused Bequest in that regard, with 8 answers that were discarded either because they were too broad or showed that the participant didn't properly comprehend the Smell.
\subsection{Middle Man}
Middle Man was revealed to be a confusing code smell for less experienced participants. Nonetheless, most of the arguments for accepting these smells directly address its basic definition: the class actually had delegated its responsibilities to another class. On the other hand, developers employed different heuristics for rejecting the code smell, including verifying the use of only local data, verifying the number of methods playing a middle man role, and even considering the incompleteness of the class evaluated.

\subsection{Primitive Obsession}
Primitive obsession has a relatively straightforward definition: several primitive variables are improperly used. In most cases, the reported code smell was accepted/rejected once the participants concluded that complex types could/could not replace the primitive variables found. However, in some cases, the participants opted by rejecting the incidence of the code smell once they concluded that some primitive elements are needed due to different reasons according to the code snippet evaluated.

\subsection{Refused Bequest}
Refused Bequest was another code smell in which developers frequently used heuristics based on its generic definition. Another heuristics adopted was checking whether inherited methods are unused or merely overridden. As Middle Man, this smell also showed to be confusing to some participants. From the 36 answers given, we discarded nine due to the lack of clear arguments.

\subsection{Speculative Generality}
Speculative Generality addresses code elements only designed for future purposes. Developers applied different heuristics for accepting the incidence of this smell. For classes, the heuristics include checking the lack of methods and the pertinence of inheritance relationships. For methods, the heuristics coded include analysing its external use and the lack of responsibilities.
    
%\subsection{Overall Analysis}

%The experiment yielded enough data so we were able to get a good understanding of how developers look at code when validating Code Smells previously detected by automated detection tools. 

%The participants criteria to define the Smells: Data Class, Feature Envy, Middle Man, Primitive Obsession and Refused Bequest, for the most part didn't deviate from the traditional definitions.
    
%A prevalent criteria for the detection of the God Class Smell was the number of lines of code, that could possibly point that developers have a tendency to look at the wrong thing when validating this particular Code Smell. Since, as previously mentioned, a high number of lines is often a byproduct of the class having many responsibilities.
    
%The mentioning of complex parameter for the Long Parameter List Smell indicates that complex parameters hinders the comprehension of methods. At first glance that would make sense, considering it's a kind of variable that nests multiple data, but needs further investigation.

%The higher number of discarded answers for Refused Bequest and Middle Man could indicate that they're harder for developers to get a proper understanding of than the rest. Nonetheless we were able to develop heuristics that should be a good complement for automated tools, increasing the effectiveness of detection when paired with them.

\section{Composing Validation Items}
\label{sec:heurstics}

Based on the findings of our study, we compiled the first version of validation items for supporting the developers in reflecting on the incidence of code smells (see Table \ref{tab:questions}). These items are expressed through questions designed to stimulate developers' rationale during the validation tasks. Considering the typical large incidence of code smells, we intentionally condensed the heuristics coded to each code smell type into a smaller set of validation items. 

Different from the metric/rule-based heuristics applied for automated detection, our validation items do not intend to be deterministic. We do not propose any conclusive thresholds nor any recommendation for accepting or rejecting code smells. However, we expect that the set of validation items proposed could help developers minding different and relevant perspectives before their final decision. Besides, it is important to note that we designed the validation items to validate candidates to code smells previously detected. Thus, we do not recommend using them as a surrogate of detection tools.

\begin{table}[ht!]
	\centering
	\caption{Validation items designed to support developers analysing the incidence of code smells.}
	\label{tab:questions}
    \begin{tabular}{|l|}
    \hline
\textbf{Data Class} \\
\hline
1-Does the class have other methods than getters and setters?\\
2-Does the class have other methods than its constructor?\\
3-Is the class data being externally manipulated?\\\\
\hline
\textbf{Feature Envy}\\\hline
1-Does the method call external methods too frequently?\\
2-Can you visualize an alternative implementation of this method focused on \\manipulating its own data?\\\\\hline

\textbf{God Class}\\\hline
1-Does the class have clear responsibilities from other classes?\\
2-Does it make sense for you to split this class into two or more classes?\\
3-Does the class size hinder its readability/comprehensibility?\\\\\hline

\textbf{Long Parameter List}\\\hline
1-Does the method signature have too many parameters?\\
2-Are there too many parameters composed of complex types?\\
3-Do the parameters' names contribute to reaching a clear understanding \\of their purpose?\\
4-Does the method actually use all its parameters?\\
5-Are all parameters actually needed?\\
6-May the parameters be passed more simply?\\\\\hline

\textbf{Middle Man}\\\hline
1-Does the class perform any relevant logical task?\\
2-Does the class clearly delegate its responsibilities to other classes?\\\\\hline

\textbf{Primitive Obsession}\\\hline
1-Does replacing one or more primitive variables with objects sound to be \\the best choice?\\
2-May two or more variables be consolidated into a single complex type?\\\\\hline

\textbf{Refused Bequest}\\\hline
1-Does the inheritance conceptually make sense?\\
2-Does the class inherit methods never used?\\
3-Does the class inherit methods that are not adherent with its definition?\\
4-Are there too many methods being overridden?\\\\\\
\hline
\end{tabular}
\end{table}

\section{Threats to Validity}
\label{sec:threats}
An important threat to the validity of our study addresses the influence of the code snippets' characteristics over the heuristics coded. In other words, if the participants evaluated another set of code snippets, probably different heuristics would be found. To mitigate this bias, we acted in two moments. In the study design, we selected code snippets from different projects having different structural characteristics to the eight types of code smell investigated. During the composing of the validation items, we tried to capture the general essence of the heuristics.

Another important threat address the researchers' bias in the coding activities. To mitigate them, the first author performed each coding step followed by a meeting with the second author for double-checking and solving disagreements. Then, both authors collaboratively worked on composing the validation items. Despite that, we are aware that some level of bias would persist. Thus, we plan to submit the validation items to the assessment of several specialists in code smells.

Finally, we are aware that the general definitions given to each code smells during the execution may have influenced their arguments. However, the results indicated that several participants frequently went beyond these definitions in their arguments. Besides, we understand that the definitions allowed the less experienced developers on smell identification to provide useful arguments.

\section{Conclusion and Future Work}
\label{sec:conclusion}
It is undeniable that developers should give the final word about the incidence of code smells. In this way, some settings on allocating developers in smells identification tasks may work better than others. However, the most common setting involves developers performing this task individually. In this paper, we propose a set of validation items for supporting the manual validation of code smells. This set is composed of questions intentionally designed to lead developers to reflect on relevant aspects of the source code according to the code smell type. These questions were depicted after coding and grouping heuristics from the arguments given by experienced developers for accepting/rejecting the incidence of code smells in several code snippets.

We are currently planning the empirical evaluation of the validation items with specialists in code smell detection. For this purpose, we are designing an opinion survey involving researchers experienced in code smells from different institutions. Through this evaluation, we intend to observe the pertinence and perceived relevance of the validation items proposed. Besides, we also intend to improve the original set of validation items by adding new heuristics proposed by these specialists.

After evolving the validation items, we plan to integrate them into an automated detection tool. With this integration, developers would set the detection tool for asking about each validation item according to the type of code smell reported. Although the validation items may come to developers' minds along the time, developers may also use the answers given to the validation items for empowering their communication during the tasks. Besides, we plan to conduct a controlled study to assess the contribution of this integrated solution over the effectiveness of the smell validation tasks.

\section{Acknowledgements}
\label{sec:acknowledgements}

We thank to the students from PUC-Rio involved in this study. We also thank to Mario Hozano and Anderson Uchoa to the valuable contributions to this work. This research was supported by PIBIC-Cefet/RJ and by CNPq 152179/2020-8.

\bibliographystyle{splncs04}
\bibliography{CIBSE_heuristics}

\end{document}